\documentclass[prb,twocolumn,floatfix,showpacs,superscriptaddress]{revtex4} 
\usepackage{graphicx,natbib,color} 
\usepackage{amsmath,amssymb}
\begin{document}

\title{Short-time dynamics of a family of XY noncollinear magnets}
\author{S. Bekhechi} 
\affiliation{Department of Physics, University of Ottawa, 
Ottawa Ontario, Canada K1N 6N5} 
\author{B.W. Southern} 
\affiliation{Department of Physics and 
Astronomy, University of Manitoba, Winnipeg Manitoba, Canada R3T 
2N2} 
\author{A. Peles} 
\affiliation{School of Physics, 
Georgia Institute of Technology, Atlanta Georgia, 30332-0430 USA} 
\author{D. Mouhanna} 
\affiliation{Laboratoire de Physique Th\'eorique de la Mati\`ere Condens\'ee, Universit\'e 
Pierre et Marie Curie, 4 Place Jussieu, 75252 Paris Cedex 05, France}
\date{\today}

\begin{abstract} 
Critical scaling and universality in the short-time dynamics for 
antiferromagnetic models on a three-dimensional stacked triangular 
lattice are investigated using Monte Carlo simulation. We have 
determined the critical point by searching for the best power law for 
the order parameter as a function of time and measured the critical 
exponents. Our results indicate that it is possible to distinguish 
weak first-order from second-order phase transitions and confirm that 
XY antiferromagnetic systems undergo a (weak) first order phase 
transition.
\end{abstract}
\maketitle

Frustrated systems\cite{Toulouse,MoesRam} are characterized by 
competing interactions which may arise due to either disorder or 
geometry. The behavior of such systems is often unpredictable but the 
basic concepts of frustrated systems may provide insights into the 
physics of complex systems and have practical uses in areas ranging 
from microelectronics to drug delivery\cite{apramirez,schiffer,shlee}. 
Magnetic systems provide simple examples of frustration where exotic 
cooperative phases such as the 'spin glass', 'spin liquid' and 'spin 
ice' are found.
During the past twenty five years a great deal of research effort has 
been put into investigating the nature of phase transitions in 
Heisenberg and XY frustrated systems in three 
dimensions.\cite{diep,collins,kawrew,NPRG,NPRG1,calabrese6} Particular 
attention has been devoted to Heisenberg and XY stacked triangular 
antiferromagnets that are commonly referred to as STA models. These 
models represent the simplest situation of frustration induced by the 
geometry of the lattice leading to a critical behavior distinct from 
that encountered in the usual ferromagnetic case. Indeed, in a 
triangular lattice, the competition due to antiferromagnetic 
interactions between nearest neighbour spins leads to a ground state 
with a planar spin configuration. In each elementary triangular cell 
the spins form a 120$^{\circ}$ structure with the vector sum of the three 
spins is equal to zero: 
\begin{equation} 
{\bf S}_A+{\bf S}_B+{\bf S}_C={\bf 0}\ 
\end{equation} 
where the subscripts $A,B,C$ label the sites at the corner of the 
elementary triangles shown in figure 1. As a consequence the order 
parameter is no longer a simple vector but a matrix, a fact that has 
lead to the idea that these (noncollinear) frustrated magnets could 
belong to a new ``chiral'' universality class\cite{kawsta}.
\begin{figure}[bp]
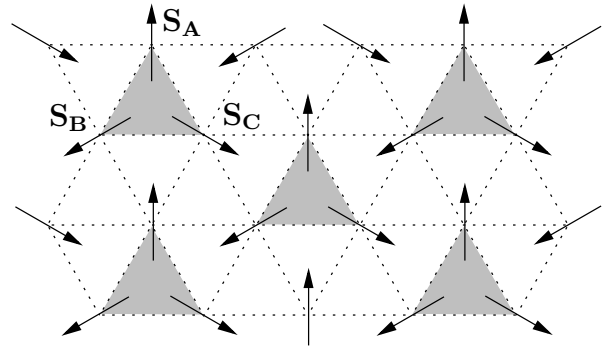
 
\centering \input plaquettes.pstex_t \caption{Ground state of the 
STA system. The shaded triangles represent elementary plaquettes.} 
\label{fig_plaquettes} 
\end{figure}

The XY STA model has been used to describe a great number of stacked 
triangular materials including CsCuCl$_3$, CsNiCl$_3$, CsMnI$_3$ and 
CsCuCl$_3$ as well as the XY Helimagnets Ho, Dy and Tb. The 
experimental results indicate that these materials exhibit second 
order phase transitions with the exception of CsCuCl$_3$ where the 
transition\cite{Weber1} is found to be weakly first order. The 
measured critical exponents exhibit scaling laws but vary from 
material to material which contradicts the basic idea of a unique set 
of critical exponents for all materials described by the same 
model. In some experiments and also in some numerical simulations, the 
critical exponent $\eta$, also called the anomalous dimension, is 
negative. This is forbidden if the theory which describes the 
transition is a unitary Landau Ginzburg Wilson (LGW) 
model\cite{justin}. Theoretical investigations using a perturbative 
renormalization group (RG) calculation up to high order predict 
the existence of a fixed point and, thus, the possibility of a 
second order phase transition\cite{calabrese6,n18}. The varying 
critical exponents in this study are associated with a spiral-like RG 
flow to a chiral, {\sl focus} fixed point\cite{n21}. Non-perturbative 
RG methods (NPRG) predict a weak first order phase transition and 
attribute the appearance of scaling by a slowing down of the RG flow 
in the whole region of the coupling constant space\cite{NPRG,NPRG1}.

The first numerical investigation of these STA models using Monte 
Carlo methods indicated a second order phase transition with set of 
critical exponents possibly associated with a new chiral universality 
class\cite{kawsta}. Some subsequent numerical investigations have been 
performed on a modified version of the STA model\cite{n19}, the STAR 
model, with the R representing a rigid constraint. In this model, the 
120$^{\circ}$ structure of the ground state is {\sl locally} imposed at 
{\sl all} temperatures. As a consequence the fluctuations of the 
spins {\sl within} a triangular cell are suppressed while the 
fluctuations in the relative orientation of the disconnected 
triangular cells, or plaquettes, can still occur. Note that the STA 
and STAR models have the same symmetries and the ``microscopic'' 
changes performed are supposed to be irrelevant to the critical 
behavior {\sl if} it is universal. In fact, it was found that the STAR 
model exhibits a strong first order phase transition thus raising 
doubts about the second order character of the phase transition 
occuring in all XY STA models. Finally, recent numerical studies of 
the STA model and its LGW formulation by Itakura\cite{itakura} also 
indicate a first order phase transition for the STA XY model itself.

In order to examine this effect of local rigidity in more detail, we 
introduced a generalized model in which we can continuously tune the 
local rigidity from the STA to the STAR limits\cite{peles,amra}, 
\begin{eqnarray} 
H(r)= -\sum_{\langle i j\rangle} J_{ij} {\bf S}_i . {\bf S}_{j} + r \sum_{plaquettes} ( 
{\bf S}_A+{\bf S}_B+{\bf S}_C)^2 
\end{eqnarray} 
The interactions $J_{ij}$ are antiferromagnetic within the triangular 
layers and ferromagnetic between layers and have the same magnitude 
$J=1$. The subscripts $A,B,C$ label the three sublattices on the 
corners of each elementary triangle and the plaquettes refer to 
disconnected triangles as shown in figure 1. The parameter $r$ imposes 
a constraint on the short wavelength fluctuations of the order 
parameter. Continuous changes in $r$ from zero to infinity correspond 
to a continuous change from the STA to the STAR model.
\begin{figure}[b] 
\centering 
\includegraphics[height=65mm,width=85mm,angle=0]{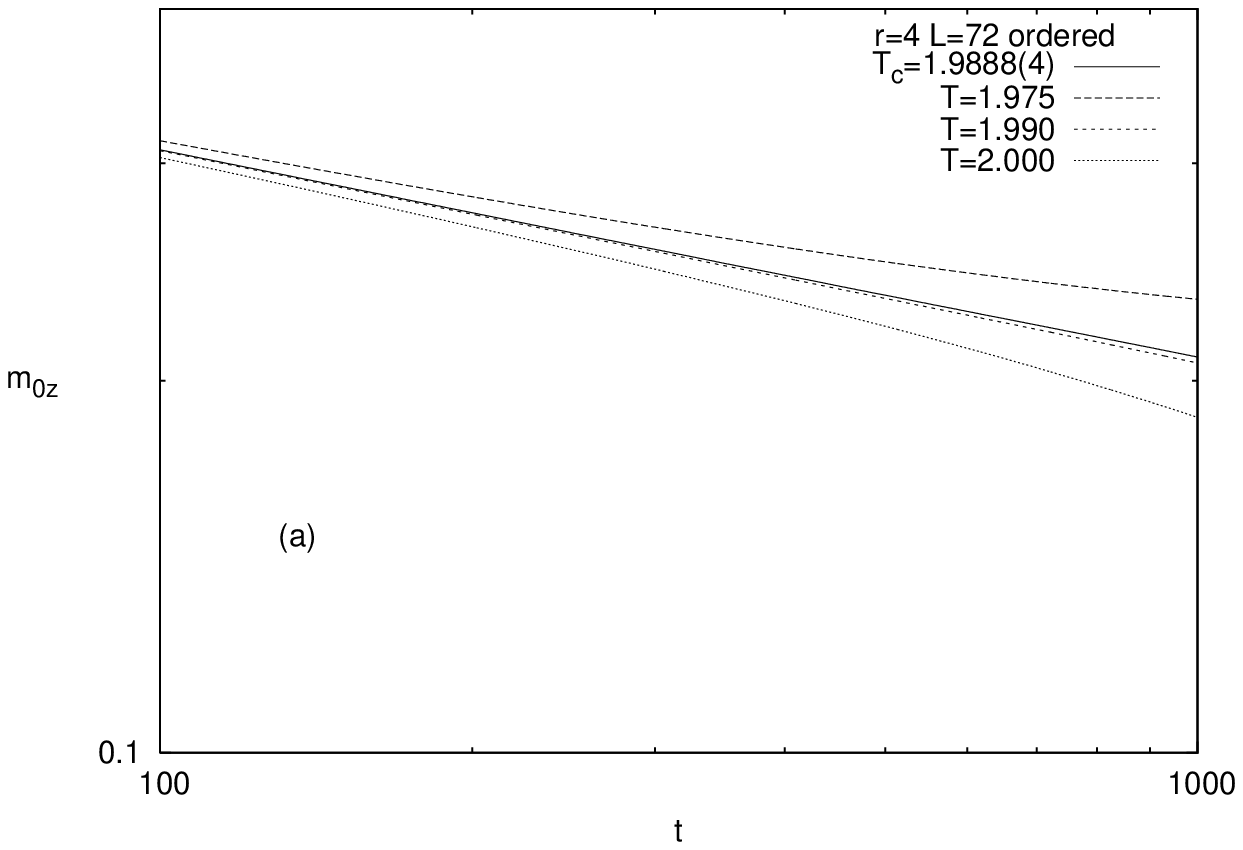} 
\includegraphics[height=65mm,width=85mm,angle=0]{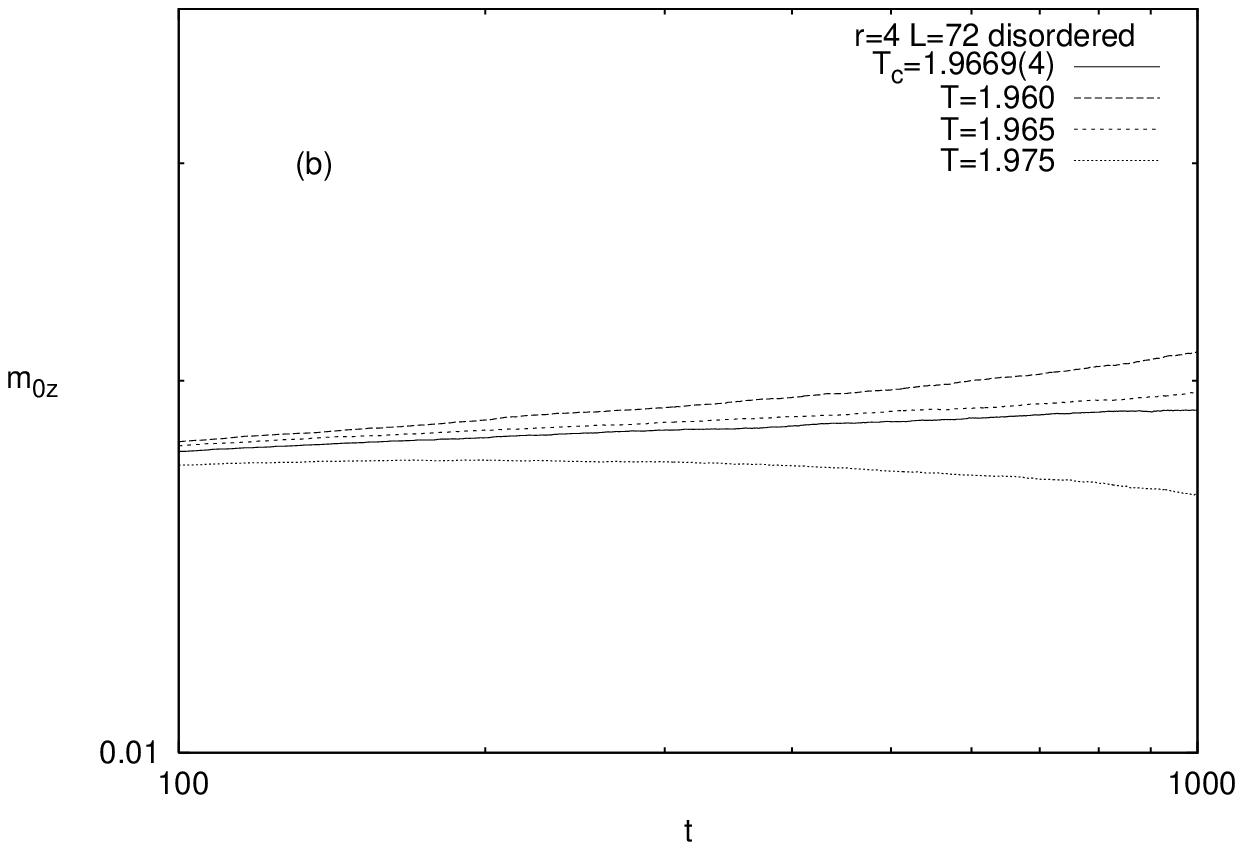}
\caption{The $z$ component of the order parameter $m_0$ as a function of 
time for $r=4$ and $L=72$ with $(a)$ an ordered initial state and 
$(b)$ a disordered initial state for various temperatures. The solid 
line is obtained by quadratic interpolation and a least square fit to 
the expected power law behavior.} 
\end{figure}

In our previous work with systems of linear sizes $L < 60$ we found 
two different types of behavior: for $r < 1.0$ the system exhibits a 
'pseudo-critical' behavior whereas, for $r >1.0$, a first order phase 
transition occurs. The critical exponents obtained in the $r < 1.0$ 
range appear to vary with the rigidity parameter $r$. This 
nonuniversal behavior is inconsistent with true critical behavior at a 
continuous phase transition for systems having the same symmetry of 
the order parameter. We concluded that the critical exponents are 
really 'pseudocritical' exponents and the observed scaling is 
'pseudoscaling'. The estimated values of critical exponents are within 
the range of the experimentally observed critical exponents for 
ABX$_3$ compounds and Tb. In the range $r > 1.0$ we were able to 
estimate the value of latent heat for several values of $r$. We 
extrapolated the values of the latent heat to $r =0$ and we found a 
small but nonzero latent heat for the XY STA model which indicated a 
very weak first order phase transition. This behavior was confirmed by 
studying the energy probability distribution using much larger systems 
sizes $L=96,138$. Even larger sizes would be needed in the case of the 
Heisenberg model\cite{itakura}. At negative values of $r$, the 
plaquettes are aligned ferromagnetically but interact 
antiferromagnetically and the symmetry of the order parameter is the 
same as at positive $r$. A special case occurs at $r =-1/2$ where we 
simply have a system of stacked Kagome layers. Additional degeneracies 
are expected in this case.
\begin{figure}[b] 
\centering 
\includegraphics[height=65mm,width=85mm,angle=0]{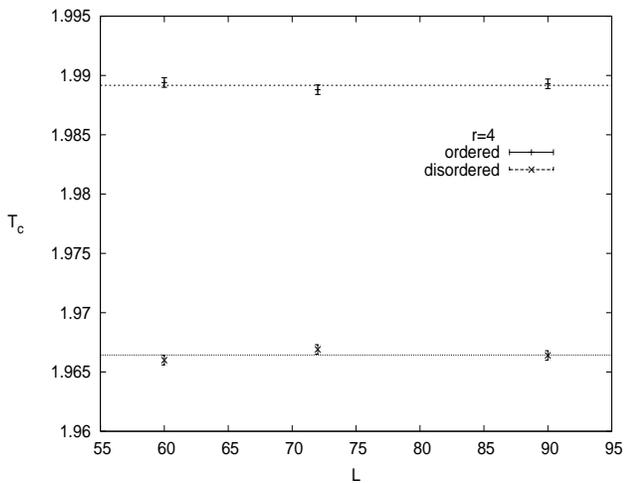} 
\caption{The $r=4$ critical temperatures for both disordered and ordered initial 
states plotted as a function of $L$. 
The lines indicate the average $T_c$ in each case and predict a $\Delta T_c 
= 0.023(1)$} 
\end{figure}

\begin{figure}[tbp] 
\centering 
\includegraphics[height=65mm,width=85mm,angle=0]{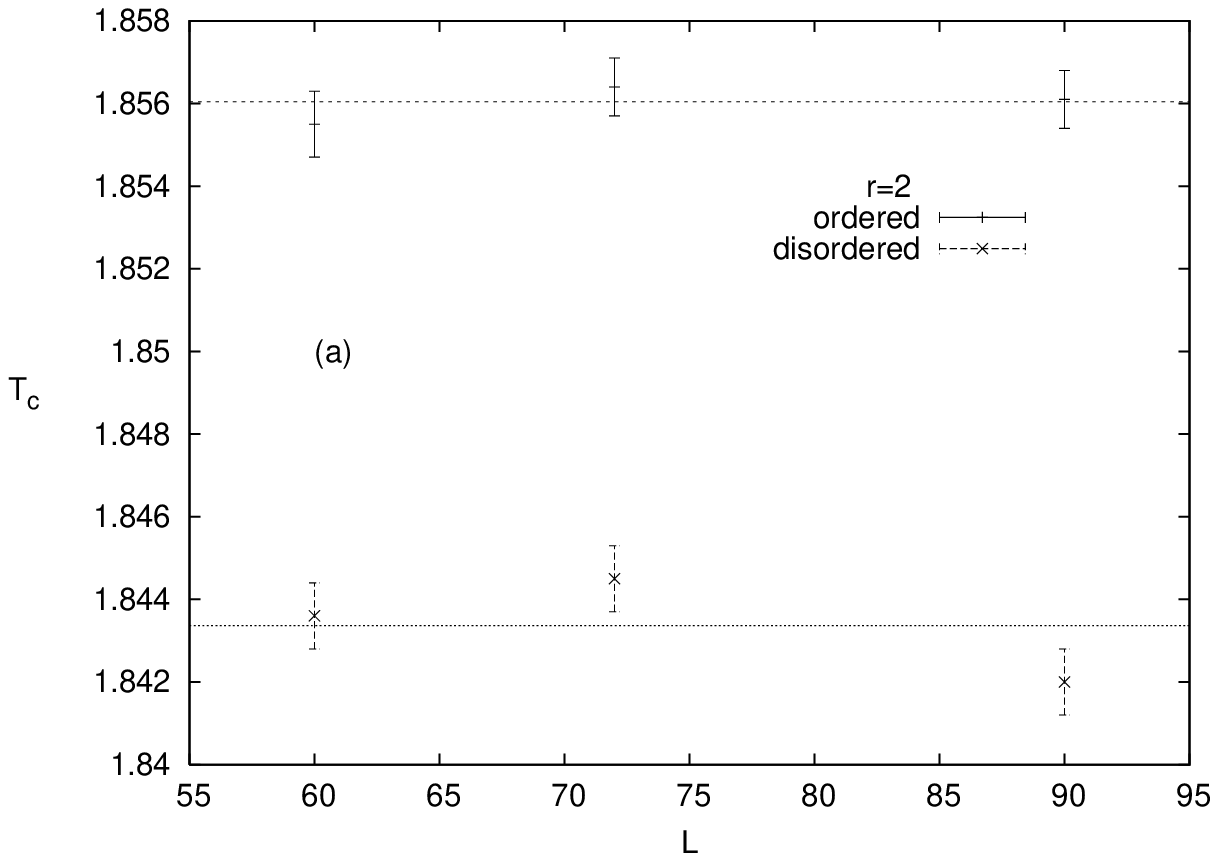} 
\includegraphics[height=65mm,width=85mm,angle=0]{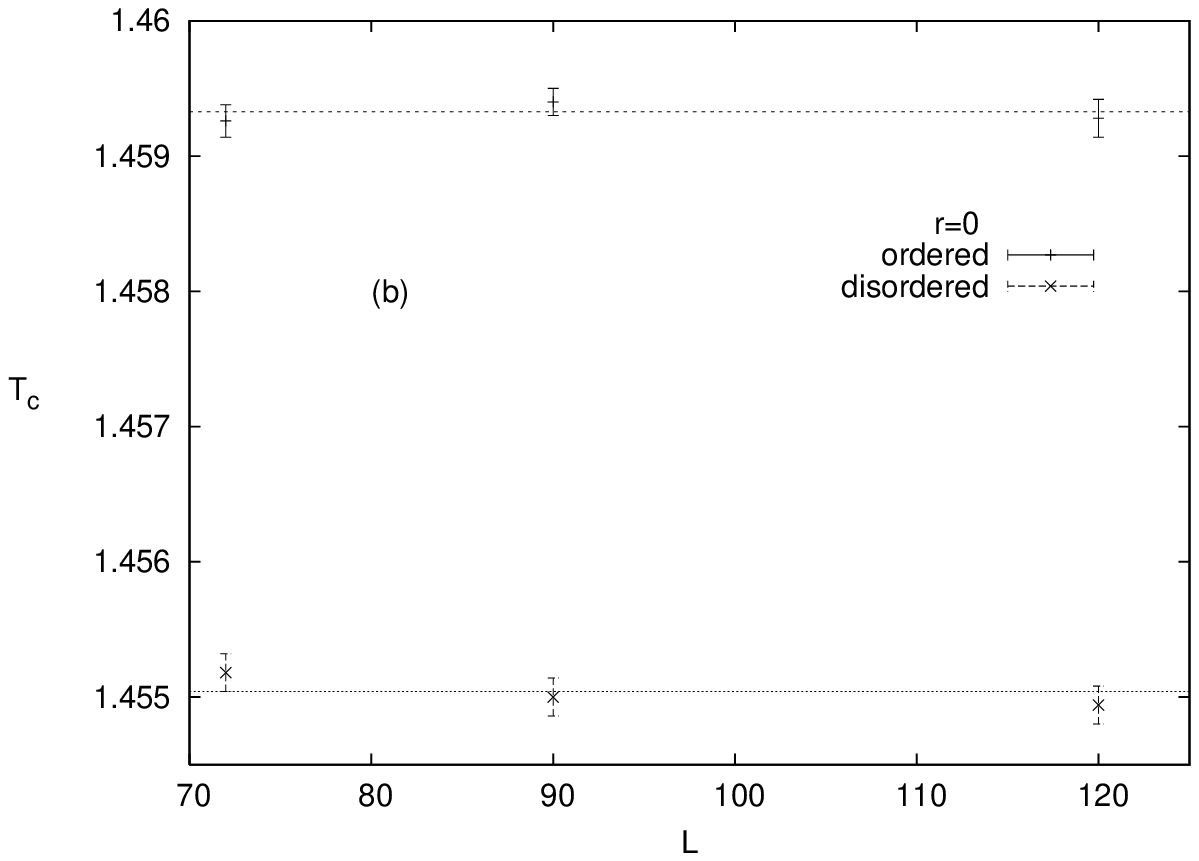}
\caption{The critical temperatures for both disordered and ordered initial states 
plotted as a function of $L$ for $(a)$ $r=2$ and $(b)$ $r=0$ . The 
lines indicate the average $T_c$ in each case and predict a $\Delta T_c 
=0.013(1) (r=2)$ and $\Delta T_c = 0.0043(1) (r=0)$ }
\end{figure} 

The standard {\it equilibrium} Monte Carlo approach requires extremely 
large lattice sizes $L$ and long runs to properly sample statistically 
independent configurations. Hence, reaching a definite conclusion 
about the nature of the phase transition that occurs in STA models 
requires a different approach. This need is even more important for 
the Heisenberg STA which, from a numerical or theoretical point of 
view, is expected to be closer to a second order behavior than the XY 
STA. For this reason we use an approach based on short time critical 
dynamics\cite{zheng,schulke}.
Universality and scaling have been observed in systems {\it far from 
equilibrium}. The initial state can be a high $T$ state which is 
rapidly quenched to the critical temperature or it can be a completely 
ordered state heated up to this temperature. Janssen 
et. al.\cite{janssen} showed that if a system is prepared at high $T$ 
with an initial value of the order parameter $m_0$ and then quenched 
to the critical temperature, the time dependent order parameter $m(t)$ 
obeys the following scaling form after a small microscopic time 
\begin{eqnarray} 
m(t,\tau,L,m_0)=b^{-\beta/\nu}m(t/b^z,b^{1/\nu}\tau,L/b,b^{x_0}m_0) 
\end{eqnarray} 
At the critical point ($\tau=0$), this form predicts that $m(t)$ 
increases as $m_0$ t$^\theta$ where $\theta=(x_0 - \beta/\nu)/z$ is a new critical 
exponent and the other exponents $\beta,\nu,z$ are the usual critical 
exponents. Higher moments of the order parameter behave similarly. The 
critical temperature $T_c$ is located by starting from an initial 
value of $m_0$ at high $T$ and then quenching to a lower $T \sim T_c$. A 
log-log plot of $m(t)$ versus time should be linear and the slope 
yields $\theta$. An ordered state can also be used as an initial state to 
estimate $T_c$ and in this case the slope yields $\beta/\nu z$ . For a 2nd 
order transition these values of $T_c$ should agree and the exponents 
should be universal. The derivative $\partial_\tau \ln m(t,\tau) |_{\tau=0}$ should 
also exhibit power law behavior with the exponent $1/\nu z$. The 
advantage of this approach is that critical slowing down is not an 
issue since the measurements are carried out at short times. Averages 
are performed over different realizations of the initial values of 
$m_0$ and thus time averages are replaced by sample averages and the 
full power of parallelized codes can be used.

The ground state of the model has the spins on the corners of each 
elementary triangle arranged at 120$^{\circ}$ to one another. There are 
three sublattices as indicated in figure 1 and the order parameter can 
be chosen as 
\begin{eqnarray} 
{\bf m}_0 ={\bf S}_A - 0.5 {\bf S}_B - 0.5 {\bf S}_C 
\end{eqnarray} 
We consider both a low temperature and high temperature initial state 
and then we follow the order parameter as a function of Monte Carlo 
time steps at some intermediate fixed temperature using the Metropolis 
algorithm. The results for each initial state are averaged over $10^3 
- 20 \times 10^3$ trials depending on the linear size $L$.

Figure 2 shows the behavior for a lattice of linear size $L=72$ and 
$r=4$ for several different temperatures starting from $(a)$ the 
ordered state and $(b)$ the disordered state. For the ordered state 
shown in figure 1 only one component of $m_0$ is non zero and we label 
this component as $m_{0z}$. For the ordered initial state it has its 
maximum value and for all disordered initial states it has the initial 
value $0.01$. We follow the time dependence of $m_{0z}$ as a function 
of time $t$ for three temperatures and we use an interpolation scheme 
to determine the temperature $T_c$ which yields the best power law 
behavior in the time range $[100,1000]$ as shown by the solid curve in 
the figures. This intermediate time range was found to give the best power law behavior
which only emerges after a
time period which is long in microscopic terms\cite{zheng}.
 The errorbars on $T_c$ are determined by the number of 
intermediate temperatures used in the interpolation scheme. We have 
performed the same calculations for sizes $L=60,72,90$ and the values 
of $T_c$ are shown in figure 3 plotted versus $L$. The values of $T_c$ 
are independent of $L$ in this time interval and the system displays 
hysteresis with the difference in the values of $T_c$ corresponding to 
the ordered and disordered inital states yielding a value of $\Delta T_c = 
0.023(1)$.
We have carried out the same procedure for smaller values of 
$r$. Figures 4(a), 4(b) show the behavior of $T_c$ for $r=2$ and $r=0$ 
respectively. In both cases the results indicate a weak first order 
transition and the values of the critical temperatures obtained here 
using the present approach straddle those obtained using equilibrium 
methods\cite{amra}. Table I summarizes the results for the 
differences in $T_c$ as determined from the ordered and disordered 
initial states as well as estimates for the various critical exponents 
obtained from the best power law dependence on $t$. The values of the 
exponents vary with the constraint parameter $r$ which indicates 
nonuniversal behavior. Using our measured values of $\beta / \nu z$ and $1/ 
\nu z$ we estimate the values of $\beta$ given in the last column. The 
values of $\beta$ increase as $r$ decreases in agreement with our previous 
study using equilibrium methods.\cite{peles,amra} The value of 
$\beta=0.27(1)$ for $r=0$ is slightly larger than that predicted by 
previous numerical studies but is consistent with the value obtained 
in experiments on STA XY antiferromagnets.\cite{NPRG}

\begin{table}[t] 
\caption{Results for the difference in critical temperatures $\Delta T_c$ and the 
exponents for various values of the constraint parameter $r$. } 
\begin{tabular}{|c|c|c|c|c|c|} 
\hline 
$r$ &$\Delta T_c$&$\theta$&$\beta/\nu z$ &$1/\nu z$&$\beta$\\ 
\hline 
\hline 
0& $0.0043(1)$ &$0.081(4)$ &$0.218(1)$&$0.79(2)$ &$0.27(1)$\\ 
\hline 
2& $0.013(1)$ &$0.045(5)$ &$0.188(2)$&$0.83(2)$ &$0.23(1)$\\ 
\hline 
4& $0.023(1)$ &$0.028(5)$ &$0.169(1)$ &$0.82(1)$ &$0.20(1)$\\ 
\hline 
\end{tabular} 
\end{table}
In summary, we have investigated the critical behavior of a family of 
XY noncollinear magnets on the stacked triangular lattice geometry 
using the short-time dynamics approach. The critical temperatures 
obtained using this approach straddle those obtained previously using 
equilibrium methods and indicate that the transition is accompanied by 
hysteresis. The critical exponents are found to vary with the 
constraint parameter $r$. Since this parameter does not change the 
symmetry of the model the exponents are non-universal. Our results 
strongly suggest that the phase transition of STA XY antiferromagnets 
is weakly first order in agreement with the NPRG field theory 
predictions and with our previous equilibrium Monte Carlo results. 
 The method used here
has the advantage that scaling behavior emerges at relatively
short times and also for smaller sizes since our values of
$T_c$ are almost independent of $L$. The results indicate 
that, for the STA XY materials, experiments need to carried out at 
reduced temperatures $\tau \ll 10^{-3}$ in order to identify the true weak 
first order nature of the transition. The present method also provides
a way to study the question of a even weaker first order transition  
for  STA Heisenberg 
materials.
\begin{acknowledgments}
This work was supported by the Natural Sciences and Research Council 
of Canada, the University of Manitoba Research Grants Program and the 
High Performance Computing facilities at the University of Manitoba 
and HPCVL Canada. We wish to thank B. Delamotte and M. Tissier for 
many useful discussions and also thank the referees for valuable 
comments and suggestions. 
\end{acknowledgments}


\begin{thebibliography}{23} 
\expandafter\ifx\csname natexlab\endcsname\relax\def\natexlab#1{#1}\fi 
\expandafter\ifx\csname bibnamefont\endcsname\relax 
\def\bibnamefont#1{#1}\fi 
\expandafter\ifx\csname bibfnamefont\endcsname\relax 
\def\bibfnamefont#1{#1}\fi 
\expandafter\ifx\csname citenamefont\endcsname\relax 
\def\citenamefont#1{#1}\fi 
\expandafter\ifx\csname url\endcsname\relax 
\def\url#1{\texttt{#1}}\fi 
\expandafter\ifx\csname urlprefix\endcsname\relax\def\urlprefix{URL }\fi 
\providecommand{\bibinfo}[2]{#2} 
\providecommand{\eprint}[2][]{\url{#2}}
\bibitem[{\citenamefont{Toulouse}(1977)}]{Toulouse} 
\bibinfo{author}{\bibfnamefont{G.}~\bibnamefont{Toulouse}}, 
\bibinfo{journal}{Commun. Phys.} \textbf{\bibinfo{volume}{{\bf 2}}}, 
\bibinfo{pages}{115} (\bibinfo{year}{1977}).
\bibitem[{\citenamefont{Moessner and Ramirez}(2006)}]{MoesRam} 
\bibinfo{author}{\bibfnamefont{R.}~\bibnamefont{Moessner}} \bibnamefont{and} 
\bibinfo{author}{\bibfnamefont{A.~P.} \bibnamefont{Ramirez}}, 
\bibinfo{journal}{Physics Today} \textbf{\bibinfo{volume}{{\bf 59 }No. 2}}, 
\bibinfo{pages}{24} (\bibinfo{year}{2006}).
\bibitem[{\citenamefont{Ramirez}(2003)}]{apramirez} 
\bibinfo{author}{\bibfnamefont{A.~P.} \bibnamefont{Ramirez}}, 
\bibinfo{journal}{Nature} \textbf{\bibinfo{volume}{{\bf 421}}}, 
\bibinfo{pages}{483} (\bibinfo{year}{2003}).
\bibitem[{\citenamefont{Schiffer}(2002)}]{schiffer} 
\bibinfo{author}{\bibfnamefont{P.}~\bibnamefont{Schiffer}}, 
\bibinfo{journal}{Nature} \textbf{\bibinfo{volume}{{\bf 420}}}, 
\bibinfo{pages}{35} (\bibinfo{year}{2002}).
\bibitem[{\citenamefont{Lee et~al.}(2002)\citenamefont{Lee, Broholm, Ratcliff, 
Gasparovic, Huang, Kim, and Cheong}}]{shlee} 
\bibinfo{author}{\bibfnamefont{S.-H.} \bibnamefont{Lee}}, 
\bibinfo{author}{\bibfnamefont{C.}~\bibnamefont{Broholm}}, 
\bibinfo{author}{\bibfnamefont{W.}~\bibnamefont{Ratcliff}}, 
\bibinfo{author}{\bibfnamefont{G.}~\bibnamefont{Gasparovic}}, 
\bibinfo{author}{\bibfnamefont{Q.}~\bibnamefont{Huang}}, 
\bibinfo{author}{\bibfnamefont{T.~H.} \bibnamefont{Kim}}, \bibnamefont{and} 
\bibinfo{author}{\bibfnamefont{S.-W.} \bibnamefont{Cheong}}, 
\bibinfo{journal}{Nature} \textbf{\bibinfo{volume}{{\bf 418}}}, 
\bibinfo{pages}{856} (\bibinfo{year}{2002}).
\bibitem[{\citenamefont{Diep}(1994)}]{diep} 
\bibinfo{editor}{\bibfnamefont{H.}~\bibnamefont{Diep}}, ed., 
\emph{\bibinfo{title}{Magnetic Systems with Competing Interactions}} 
(\bibinfo{publisher}{World Scientific, Singapore}, \bibinfo{year}{1994}).
\bibitem[{\citenamefont{Collins and Petrenko}(1997)}]{collins} 
\bibinfo{author}{\bibfnamefont{M.~F.} \bibnamefont{Collins}} \bibnamefont{and} 
\bibinfo{author}{\bibfnamefont{O.~A.} \bibnamefont{Petrenko}}, 
\bibinfo{journal}{Can. J. Phys.} \textbf{\bibinfo{volume}{{\bf 75}}}, 
\bibinfo{pages}{605} (\bibinfo{year}{1997}).
\bibitem[{\citenamefont{Kawamura}(1998)}]{kawrew} 
\bibinfo{author}{\bibfnamefont{H.}~\bibnamefont{Kawamura}}, 
\bibinfo{journal}{J. Phys.: Condens. Matter} \textbf{\bibinfo{volume}{{\bf 
10}}}, \bibinfo{pages}{4707} (\bibinfo{year}{1998}).
\bibitem[{\citenamefont{Delamotte 
et~al.}(2004{\natexlab{a}})\citenamefont{Delamotte, Mouhanna, and 
Tissier}}]{NPRG} 
\bibinfo{author}{\bibfnamefont{B.}~\bibnamefont{Delamotte}}, 
\bibinfo{author}{\bibfnamefont{D.}~\bibnamefont{Mouhanna}}, \bibnamefont{and} 
\bibinfo{author}{\bibfnamefont{M.}~\bibnamefont{Tissier}}, 
\bibinfo{journal}{Phys. Rev. B} \textbf{\bibinfo{volume}{{\bf 69}}}, 
\bibinfo{pages}{134413} (\bibinfo{year}{2004}{\natexlab{a}}).
\bibitem[{\citenamefont{Delamotte 
et~al.}(2004{\natexlab{b}})\citenamefont{Delamotte, Mouhanna, and 
Tissier}}]{NPRG1} 
\bibinfo{author}{\bibfnamefont{B.}~\bibnamefont{Delamotte}}, 
\bibinfo{author}{\bibfnamefont{D.}~\bibnamefont{Mouhanna}}, \bibnamefont{and} 
\bibinfo{author}{\bibfnamefont{M.}~\bibnamefont{Tissier}}, 
\bibinfo{journal}{J. Phys.: Condens. Matter} \textbf{\bibinfo{volume}{{\bf 
16}}}, \bibinfo{pages}{S883} (\bibinfo{year}{2004}{\natexlab{b}}).
\bibitem[{\citenamefont{Calabrese et~al.}(2004)\citenamefont{Calabrese, 
Parruccini, Pelissetto, and Vicari}}]{calabrese6} 
\bibinfo{author}{\bibfnamefont{P.}~\bibnamefont{Calabrese}}, 
\bibinfo{author}{\bibfnamefont{P.}~\bibnamefont{Parruccini}}, 
\bibinfo{author}{\bibfnamefont{A.}~\bibnamefont{Pelissetto}}, 
\bibnamefont{and} \bibinfo{author}{\bibfnamefont{E.}~\bibnamefont{Vicari}}, 
\bibinfo{journal}{Phys. Rev. B} \textbf{\bibinfo{volume}{{\bf 70}}}, 
\bibinfo{pages}{174439} (\bibinfo{year}{2004}).
\bibitem[{\citenamefont{Kawamura}(1992)}]{kawsta} 
\bibinfo{author}{\bibfnamefont{H.}~\bibnamefont{Kawamura}}, 
\bibinfo{journal}{J. Phys. Soc. Japan} \textbf{\bibinfo{volume}{{\bf 61}}}, 
\bibinfo{pages}{1299} (\bibinfo{year}{1992}).
\bibitem[{\citenamefont{Weber et~al.}(1996)\citenamefont{Weber, Werner, 
Wosnitza, v.~L{\"{o}}hneysen, and Schotte}}]{Weber1} 
\bibinfo{author}{\bibfnamefont{H.~B.} \bibnamefont{Weber}}, 
\bibinfo{author}{\bibfnamefont{T.}~\bibnamefont{Werner}}, 
\bibinfo{author}{\bibfnamefont{J.}~\bibnamefont{Wosnitza}}, 
\bibinfo{author}{\bibfnamefont{H.}~\bibnamefont{v.~L{\"{o}}hneysen}}, 
\bibnamefont{and} \bibinfo{author}{\bibfnamefont{U.}~\bibnamefont{Schotte}}, 
\bibinfo{journal}{Phys. Rev. B} \textbf{\bibinfo{volume}{{\bf 54}}}, 
\bibinfo{pages}{15924} (\bibinfo{year}{1996}).
\bibitem[{\citenamefont{Zinn-Justin}(1989)}]{justin} 
\bibinfo{author}{\bibfnamefont{J.}~\bibnamefont{Zinn-Justin}}, 
\emph{\bibinfo{title}{Quantum Field Theory and Critical Phenomena}} 
(\bibinfo{publisher}{Oxford University Press, New York, 3rd ed.}, 
\bibinfo{year}{1989}).
\bibitem[{\citenamefont{Pelissetto et~al.}(2001)\citenamefont{Pelissetto, 
Rossi, and Vicari}}]{n18} 
\bibinfo{author}{\bibfnamefont{A.}~\bibnamefont{Pelissetto}}, 
\bibinfo{author}{\bibfnamefont{P.}~\bibnamefont{Rossi}}, \bibnamefont{and} 
\bibinfo{author}{\bibfnamefont{E.}~\bibnamefont{Vicari}}, 
\bibinfo{journal}{Phys.\ Rev. B} \textbf{\bibinfo{volume}{{ \bf 63}}}, 
\bibinfo{pages}{140414(R)} (\bibinfo{year}{2001}).
\bibitem[{\citenamefont{Calabrese et~al.}(2002)\citenamefont{Calabrese, 
Parruccini, and Sokolov}}]{n21} 
\bibinfo{author}{\bibfnamefont{P.}~\bibnamefont{Calabrese}}, 
\bibinfo{author}{\bibfnamefont{P.}~\bibnamefont{Parruccini}}, 
\bibnamefont{and} \bibinfo{author}{\bibfnamefont{A.~I.} 
\bibnamefont{Sokolov}}, \bibinfo{journal}{Phys. Rev. B} 
\textbf{\bibinfo{volume}{{\bf 66}}}, \bibinfo{pages}{180403R} 
(\bibinfo{year}{2002}).
\bibitem[{\citenamefont{Loison and Schotte}(2000)}]{n19} 
\bibinfo{author}{\bibfnamefont{D.}~\bibnamefont{Loison}} \bibnamefont{and} 
\bibinfo{author}{\bibfnamefont{K.~D.} \bibnamefont{Schotte}}, 
\bibinfo{journal}{Eur. Phys. J. B} \textbf{\bibinfo{volume}{{ \bf 14}}}, 
\bibinfo{pages}{125} (\bibinfo{year}{2000}).
\bibitem[{\citenamefont{Itakura}(2003)}]{itakura} 
\bibinfo{author}{\bibfnamefont{M.}~\bibnamefont{Itakura}}, \bibinfo{journal}{J. 
Phys. Soc. Jpn.} \textbf{\bibinfo{volume}{{\bf 72}}}, \bibinfo{pages}{74} 
(\bibinfo{year}{2003}).
\bibitem[{\citenamefont{Peles et~al.}(2004)\citenamefont{Peles, Southern, 
Delamotte, Mouhanna, and Tissier}}]{peles} 
\bibinfo{author}{\bibfnamefont{A.}~\bibnamefont{Peles}}, 
\bibinfo{author}{\bibfnamefont{B.~W.} \bibnamefont{Southern}}, 
\bibinfo{author}{\bibfnamefont{B.}~\bibnamefont{Delamotte}}, 
\bibinfo{author}{\bibfnamefont{D.}~\bibnamefont{Mouhanna}}, \bibnamefont{and} 
\bibinfo{author}{\bibfnamefont{M.}~\bibnamefont{Tissier}}, 
\bibinfo{journal}{Phys. Rev. B} \textbf{\bibinfo{volume}{{\bf 69}}}, 
\bibinfo{pages}{220408(R)} (\bibinfo{year}{2004}).
\bibitem[{\citenamefont{Peles}(2004)}]{amra} 
\bibinfo{author}{\bibfnamefont{A.}~\bibnamefont{Peles}}, Ph.D. thesis, 
\bibinfo{school}{University of Manitoba, Winnipeg Manitoba, Canada} 
(\bibinfo{year}{2004}).
\bibitem[{\citenamefont{Zheng}(1998)}]{zheng} 
\bibinfo{author}{\bibfnamefont{B.}~\bibnamefont{Zheng}}, \bibinfo{journal}{Int. 
J. Mod. Phys. B} \textbf{\bibinfo{volume}{{\bf 12}}}, \bibinfo{pages}{1419} 
(\bibinfo{year}{1998}).
\bibitem[{\citenamefont{Sch{\"{u}}lke and Zheng}(2000)}]{schulke} 
\bibinfo{author}{\bibfnamefont{L.}~\bibnamefont{Sch{\"{u}}lke}} 
\bibnamefont{and} \bibinfo{author}{\bibfnamefont{B.}~\bibnamefont{Zheng}}, 
\bibinfo{journal}{Phys. Rev. E} \textbf{\bibinfo{volume}{{\bf 62}}}, 
\bibinfo{pages}{7482} (\bibinfo{year}{2000}).
\bibitem[{\citenamefont{Janssen et~al.}(1989)\citenamefont{Janssen, Schaub, and 
Schmittmann}}]{janssen} 
\bibinfo{author}{\bibfnamefont{H.~K.} \bibnamefont{Janssen}}, 
\bibinfo{author}{\bibfnamefont{B.}~\bibnamefont{Schaub}}, \bibnamefont{and} 
\bibinfo{author}{\bibfnamefont{B.}~\bibnamefont{Schmittmann}}, 
\bibinfo{journal}{Zeitschrift f{\"{u}}r Physik B} 
\textbf{\bibinfo{volume}{{\bf 73}}}, \bibinfo{pages}{539} 
(\bibinfo{year}{1989}).
\end{thebibliography}
\end{document}